\documentclass[prl,twocolumn,tightenlines,preprintnumbers,floatfix,nofootinbib]{revtex4}
\usepackage{graphicx,amsfonts,amssymb,amsmath}

\def\be{\begin{equation}}
\def\ee{\end{equation}}
\def\ba{\begin{eqnarray}}
\def\ea{\end{eqnarray}}
\def\ge{\mathrel{\raise.3ex\hbox{$>$\kern-.75em\lower1ex\hbox{$\sim$}}}}
\def\la{\mathrel{\raise.3ex\hbox{$<$\kern-.75em\lower1ex\hbox{$\sim$}}}}

\def\simgt{\mathrel{\raise.3ex\hbox{$>$\kern-.75em\lower1ex\hbox{$\sim$}}}}
\def\simlt{\mathrel{\raise.3ex\hbox{$<$\kern-.75em\lower1ex\hbox{$\sim$}}}}

\newcommand{\bi}[1]{\bibitem{#1}}
\newcommand{\fr}[2]{\frac{#1}{#2}}

\newcommand{\nc}{\newcommand}

\nc{\gone}{\bar g_{\pi NN}^{(1)}}
\nc{\gzero}{\bar g_{\pi NN}^{(0)}}
\nc{\al}{\alpha}
\nc{\ga}{\gamma}
\nc{\de}{\delta}
\nc{\ep}{\epsilon}
\nc{\ze}{\zeta}
\nc{\et}{\eta}

\nc{\Th}{\Theta}
\nc{\ka}{\kappa}
\nc{\rh}{\rho}
\nc{\si}{\sigma}
\nc{\ta}{\tau}
\nc{\up}{\upsilon}
\nc{\ph}{\phi}
\nc{\ch}{\chi}
\nc{\ps}{\psi}
\nc{\om}{\omega}
\nc{\Ga}{\Gamma}
\nc{\De}{\Delta}
\nc{\La}{\Lambda}
\nc{\Si}{\Sigma}
\nc{\Up}{\Upsilon}
\nc{\Ph}{\Phi}
\nc{\Ps}{\Psi}
\nc{\Om}{\Omega}
\nc{\ptl}{\partial}
\nc{\del}{\nabla}
\nc{\ov}{\overline}
\nc{\newcaption}[1]{\centerline{\parbox{15cm}{\caption{#1}}}}
\nc{\hef}{$^4$He}
\nc{\het}{$^3$He}
\nc{\lisx}{$^6$Li}
\nc{\lisv}{$^7$Li}
\nc{\bes}{$^7$Be}
\nc{\beet}{$^8$Be}
\nc{\hefm}{^4{\rm He}}
\nc{\hetm}{^3{\rm He}}
\nc{\lisxm}{^6{\rm Li}}
\nc{\lisvm}{^7{\rm Li}}
\nc{\besm}{^7{\rm Be}}
\nc{\beetm}{^8{\rm Be}}
\nc{\bs}{(N$X^-$)}
\nc{\xm}{$X^-$}

\def\beq{\begin{equation}}
\def\eeq{\end{equation}}
\def\bmat{\begin{displaymath}}
\def\emat{\end{displaymath}}
\def\bear{\begin{eqnarray}}
\def\eear{\end{eqnarray}}
\def\bery{\begin{array}}
\def\ery{\end{array}}
\def\bit{\begin{itemize}}
\def\eit{\end{itemize}}
\def\ben{\begin{enumerate}}
\def\een{\end{enumerate}}
\def\btab{\begin{tabular}}
\def\etab{\end{tabular}}
\def\btbl{\begin{table}}
\def\etbl{\end{table}}
\def\bfig{\begin{figure}[htb]}
\def\efig{\end{figure}}
\def\bpic{\begin{picture}}
\def\epic{\end{picture}}

\def\ga{\mathrel{\raise.3ex\hbox{$>$\kern-.75em\lower1ex\hbox{$\sim$}}}}
\def\la{\mathrel{\raise.3ex\hbox{$<$\kern-.75em\lower1ex\hbox{$\sim$}}}}
\def\gappeq{\mathrel{\rlap {\raise.5ex\hbox{$>$}}
{\lower.5ex\hbox{$\sim$}}}}
\def\lappeq{\mathrel{\rlap{\raise.5ex\hbox{$<$}}
{\lower.5ex\hbox{$\sim$}}}}

\def\gyr{{\rm \, G\kern-0.125em yr}}
\def\mev{{\rm \, Me\kern-0.125em V}}
\def\gev{{\rm \, Ge\kern-0.125em V}}
\def\tev{{\rm \, Te\kern-0.125em V}}

\begin{document}


\setcounter{page}{1}


\title{Particle physics catalysis of thermal Big Bang Nucleosynthesis}

\author{Maxim Pospelov$^{\,(a,b)}$}

\affiliation{$^{\,(a)}${\it Perimeter Institute for Theoretical Physics, Waterloo,
Ontario N2J 2W9, Canada}\\
$^{\,(b)}${\it Department of Physics and Astronomy, University of Victoria, 
     Victoria, BC, V8P 1A1 Canada}
}

\begin{abstract}

We point out that the existence of metastable, $\tau>10^3$ s, negatively charged 
electroweak-scale particles (\xm) alters the predictions for lithium and other 
primordial elemental abundances for $A>4$ via the formation of 
bound states with nuclei during BBN. In particular, we show that the 
bound states of $X^-$ with helium, formed at temperatures of about $T=10^8$K, 
lead to the catalytic enhancement of $^6$Li production, which is eight orders of magnitude 
more efficient than the standard channel. In particle physics models where subsequent decay 
of \xm\ does not lead to large non-thermal BBN effects, this directly translates to the level of sensitivity 
to the number density of long-lived \xm\, particles  ($\tau>10^5$ s) relative to entropy
of $n_{X^-}/s \la 3\times 10^{-17}$, which is one of the most stringent probes of electroweak scale
remnants known to date.

\end{abstract}

\maketitle

\newpage

Standard Big Bang Nucleosynthesis (SBBN) is a well-established theory that makes predictions 
for elemental abundances of light elements, H, D, He and Li, as functions of only one 
free parameter, the ratio of baryon to photon number densities.
Agreement of the observed abundances 
for D and \hef\, with the SBBN predictions that use an additional 
CMB-derived \cite{WMAP} input value of $n_b/s=0.9\times 10^{-10}$ 
serves as a sensitive probe of New Physics. 
Indeed, over the years a finessed SBBN approach
has led to constraints on several non-standard scenarios, including constraints on
the total number of thermally excited relativistic degrees of freedom 
\cite{Sarkar}. This first class of 
constraints results essentially
from the destortion of the time-frame for the $n/p$ freeze-out, affecting mostly the abundance of \hef.
The second class of constraints comes from the nonthermal BBN, that results from 
the late decay of metastable heavy particles \cite{metastable}.
If such decays happen during or after BBN, they trigger electromagnetic cascades that affect
light elements via photodissociation, or lead to non-thermal nuclear reactions by 
fast hadrons produced in the decay,
resulting in significant modifications of 
D, \lisx\, \lisv\, and \het/D abundances.
The BBN constraints have been instrumental in limiting some variants of supersymmetric (SUSY) models with 
long-lived unstable particles. For example, late decays  put constraints on the energy density of unstable 
gravitinos, limiting it to be less than $\sim 10^{-13}$ relative to entropy$\times$GeV \cite{cefo,kkm,jj}
for some selected range of lifetimes and SM branching ratios.  
It is important to establish whether the  late decay of 
heavy particles 
can possibly ``cure" \cite{cefo,kkm,jj} what is known as the lithium problem, a statistically significant 
and persistent discrepancy of the SBBN prediction for \lisv~ from the roughly twice smaller 
observational value over a wide range of metallicities. 

The purpose of this Letter is to show that in addition to the change of the timing for BBN reaction and non-thermal processes, 
there is a third way particle physics can significantly affect the prediction for primordial 
abundances of light elements. This new way consists of the {\em catalysis} of thermal nuclear reactions 
by heavy relic particles that have long-range (electromagnetic or strong) interactions with nuclei. 
In particular, we show that the cosmological presence of metastable charged particles,
called $X^-$ hereafter enables the catalyzed BBN (CBBN) via the formation of bound states between 
light nuclei and negatively charged particles $X^-$.  These bound states form in the range of temperatures 
from 1 to 30 KeV, changing the standard nuclear reaction rates but more importantly opening 
new channels for thermal reactions and changing the abundance of Li and other elements with $A>4$. 
The most significant difference is seen in the \lisx~ production mechanism, 
\begin{eqnarray}
\label{SBBN}
&&\!\!\!\!\!\!\!\!\!\!\!\!\!\!\!\!{\rm SBBN}:~\hefm +{\rm D}\to ~\lisxm +\gamma; ~~~~~Q= 1.47 {\rm MeV}
\\
&&\!\!\!\!\!\!\!\!\!\!\!\!\!\!\!\!{\rm CBBN}: ~ (\hefm X^-) +{\rm D}\to ~\lisxm +X^-; ~Q \simeq 1.13{\rm MeV},
\label{CBBN}
\end{eqnarray}
and, as we are going to show, the cross section for the CBBN channel is enhanced by eight orders of magnitude
relative to SBBN. In (\ref{CBBN}) and below, $(NX)$
denotes the bound state of a nucleus $N$ and $X^-$. In the remainder of this Letter, 
after a brief review of the relevant bound states $(NX)$, we analyze the cross section and thermal rate
for reaction (\ref{CBBN}), make a prediction for \lisx\ in CBBN, 
put constraints on some of the particle physics scenarios, and 
point out new CBBN mechanisms for \lisv\, depletion.

{\em Properties of the bound states}. Here we assume that the electromagnetic force is acting between 
nuclei and $X^-$ particles, and calculate properties of the 
ground states using the variational approach. 
Binding energies $E_b$, average distances 
between the center of the nucleus and $X^-$, and the ``photo-dissociation decoupling" temperatures 
$T_0$, are summarized in Table I where the constraint $m_{X^-}\gg m_N$ is 
also imposed.
\begin{table}
\begin{center}
\begin{tabular}{|c|c|c|c|c|c|c|c|}
\hline
bound st. &     $|E_b^0|$&  $a_0$ & $R_N^{sc}$ & $|E_b(R_N^{sc})|$ & $R_{Nc}$  & $|E_b(R_{Nc})|$ &~$T_0$~ \\ \hline\hline
\hef$X^-$&          397  &   3.63 &    1.94    &       352         &    2.16   &         346     &  8.2 \\ \hline
\het$X^-$&          299  &   4.81 &    1.76    &       276         &    2.50   &         267     &  6.3 \\ \hline
\lisv$X^-$&        1566  &   1.38 &    2.33    &       990         &    3.09   &         870     &  21  \\ \hline
\bes$X^-$&         2787  &   1.03 &    2.33    &       1540        &      3    &         1350    &  32  \\ \hline
\beet$X^-$&        3178  &   0.91 &    2.44    &       1600        &      3    &         1430    &  34  \\ \hline
T$X^-$&              75  &    9.6  &    2.27   &       73         &     2.27   &           73    & 1.8 \\ \hline
D$X^-$&              50  &    14  &      -     &       49          &    2.75   &           49    & 1.2  \\ \hline
p$X^-$&              25  &    29  &      -     &       25          &      1.10 &           25    & 0.6  \\ \hline
\end{tabular}
\end{center}
\caption{Properties of the bound states: Bohr $a_0$ and nuclear radii $R_{N}$ in fm; binding energies $E_b$ and 
``photo-dissociation decoupling" temperatures
$ T_0$ in KeV.
}
\vspace{-0.5cm}
\label{table1}
\end{table}
 The main uncertainty in Table I comes from the charge distribution 
inside the nucleus as the naive Bohr orbit
$a_0=(Z\alpha m_N)^{-1}$ can be well within the nuclear radius.
 It leads to a reduction of the bound state energies relative to 
the Bohr-like formula, $E_b^0 = Z^2\alpha^2m_N/2$ from $\sim 13\%$ in (\hef$X$)
to 50\% in (\beet$X$). Realistic binding energies are calculated for two 
types of nuclear radii assuming a uniform charge distribution: for the simplest
scaling formula $R_N^{sc}= 1.22A^{\fr{1}{3}}$, and for the nuclear radius determined via the 
the root mean square charge radius, $R_{Nc} = (5/3)^{1/3}R_c$ with experimental input for $R_c$ where available.
Finally, as an indication of the temperature at which $(NX)$ are no longer ionized, 
we include a scale $T_0$ where the 
photo-dissociation rate $\Gamma_{\rm ph}(T)$ becomes smaller than the Hubble rate, $\Gamma_{\rm ph}(T_0)=H(T_0)$. 
It is remarkable  that stable bound states of $(\beetm X)$ exist,
opening up a path to synthesize heavier elements such as carbon, which is not 
produced in SBBN. In addition to atomic states, there exist molecular bound states $(NXX)$. 
The binding energy of such molecules relative to $(NX)$ 
are not small ({\em e.g.} about 300 KeV for (\hef\xm\xm)). Such neutral molecules, along
with (\beet$X$) and  (\beet$XX$), are an important path for the synthesis of heavier elements in CBBN. 
One can easily generalize Table 1 for the case of doubly-charged particles,
which was recently discussed in \cite{khlopov} in connection with the dark matter problem.
\begin{figure}[htbp]
\centerline{\includegraphics[width=8.2cm]{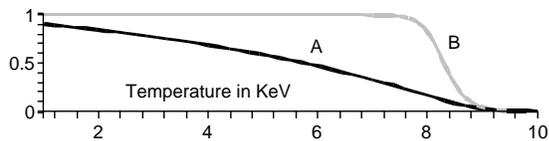}}
 \caption{\footnotesize Fraction of \xm\, locked in the bound state with \hef: 
A. Realistic result based on Boltzmann equation, B. Saha-type prediction with a rapid switch from 0 to 1 at $T\simeq 8.3$ KeV}
\label{f1} 
\end{figure}

The initial abundance of $X^-$ particles relative to baryons, 
$Y_{X}(t\ll \tau)\equiv n_{X^-}/n_b$, along with their lifetime $\tau$  
are the input parameters of CBBN, and it is safe to assume that $Y_X\ll1$. 
The most important catalytic enhancement results from the bound states ($\hefm X$). 
Their abundance relative to the total abundance of \xm, $Y_{BS}=n_{BS}/n_{X^-}$ is
calculated using the Boltzmann equation 
\be
-HT\fr{d Y_{BS}}{dT} = \langle \sigma_{rec}v\rangle n_{\rm He} - \langle \sigma_{ph}v\rangle Y_{\rm BS}n_\gamma
\label{rec}
\ee
along with the photoionization and recombination cross section.
The result is shown in Fig. 1, where a significant deviation from the naive Saha equation is observed 
as the recombination rate of $X^-$ and \hef\, is only marginally larger than the Hubble rate. For the same reason 
the abundance of bound states with rare light elements is very low,  
$n_{N X^-}/n_{X^-} \la 10^{-6}$, where $N=$\het, D and T. As we are going to show, 
an actual abundance of \xm\, at $T\simeq 8 $ KeV has to be less than $10^{-6}$, which makes the 
overall impact of these bounds states on BBN negligible.


\begin{figure}[htbp]
\centerline{\includegraphics[width=8.2cm]{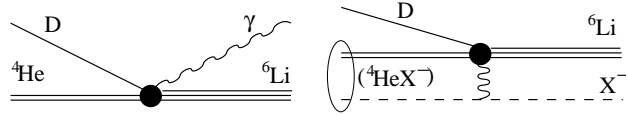}}
 \caption{\footnotesize SBBN and CBBN mechanisms for producing \lisx.}
\label{f1} 
\end{figure}

{\em Photonless production of {\rm\lisx}.} 
The standard mechanism for \lisx\, production in SBBN is ``accidentally" suppressed. 
The D-\hef\, cluster description 
gives a good approximation to this process,  and 
the reaction rate of (\ref{SBBN}) is dominated by the E2 amplitude because the 
E1 amplitude nearly vanishes due to an (almost) identical charge to mass ratio for
D and \hef. In the E2 transition, the quadrupole moment of D-\hef\,  interacts with the gradient of 
the external electromagnetic field, $V_{\rm int} = Q_{ij} \nabla_i E_j$. 
Consequently,  the cross section at BBN energies scales as the inverse fifth power of 
photon wavelength $\lambda = \omega^{-1} \sim 130$ fm, which is significantly larger than the nuclear 
distances that saturate the matrix element of $Q_{ij}$, leading to strong suppression of (\ref{SBBN}) relative to 
other BBN cross sections \cite{kawano}. 
For the CBBN process (\ref{CBBN}) the real photon in the final state
is replaced by a virtual photon with a characteristic wavelength on the order of the Bohr radius in  (\hef\xm). 
Correspondingly, one expects the enhancement factor 
in the ratio of CBBN to SBBN cross sections  to scale as $(a_0\omega)^{-5}\sim 5\times 10^7$.
Figure 1 presents a schematic depiction of both processes. It is helpful that in the limit 
of $R_N\ll a_0$, we can apply factorization, calculate the effective $\nabla_i E_j$ created by \xm, 
and relate SBBN and CBBN cross sections {\em without} 
explicitly calculating the $\langle {\rm D}\hefm|Q_{ij}|\lisxm\rangle$ matrix element. 
A straightforward quantum-mechanical calculation with $\nabla_i E_j$  averaged over 
the Hydrogen-like initial state of ($\hefm X$) and the 
plane wave of \lisx\ in the final state leads to the following relation between 
the astrophysical $S$-factors at low energy:
\be
S_{\rm CBBN} = S_{\rm SBBN} \times \fr{8}{3\pi^2}\fr{p_fa_0}{(\omega a_0)^5}\left(1+\fr{m_{\rm D}}{m_{\hefm}}\right)^2.
\label{Snaive}
\ee
Here $a_0$ is the Bohr radius of (\hef\xm), $p_f = (2m_{\lisxm}(Q_{\rm CBBN}+E))^{1/2}$ 
is the momentum of the outgoing \lisx\ in the CBBN reaction, and $\omega$ is the photon energy in the SBBN process,
$\omega = Q_{SBBN} +E$. For $E\ll Q$ the value of the final momentum of the \lisx\, 
nucleus is $p_f \simeq (1.8{\rm fm})^{-1}$. Throughout the whole paper, $c=\hbar=1$. 
The $S$ factor is defined in the standard way, by removing the 
Gamow factor $G$ from the cross section: $S(E)=E\sigma/G$. A somewhat 
more sophisticated approach with the use of an optimized 
wavefunction for (\hef$X$) in the initial state and the Coulomb wavefunction in the final state,
with the radial integrals calculated with the constraint $|{\bf r}_{X} - {\bf r}_{\lisxm}|> 2$ fm, renders the result,
which is very close to a naive estimate:
\be
S_{\rm CBBN}(0) \simeq  6\times 10^7\times 
\fr{S_{\rm SBBN}(E_0)}{(1+ E_0/Q_{\rm SBBN})^5}  \simeq 0.3~{\rm MeVbn}.
\label{Snumeric}
\ee
Here the SBBN value of $S_{\rm SBBN} = 1.8\times 10^{-2}$eVbn is taken at some matching 
scale $E_0=0.41$ MeV \cite{sgamma}, which is sufficiently close to 0, yet is in the regime
where SBBN cross sections can be reliably measured. The result of the factorized 
approach (\ref{Snumeric}) should be within a factor of a few from the exact answer, 
with the largest errors presumably associated with the $R_N/a_0$ expansion. 
\begin{figure}[htbp]
\centerline{\includegraphics[width=8.8cm]{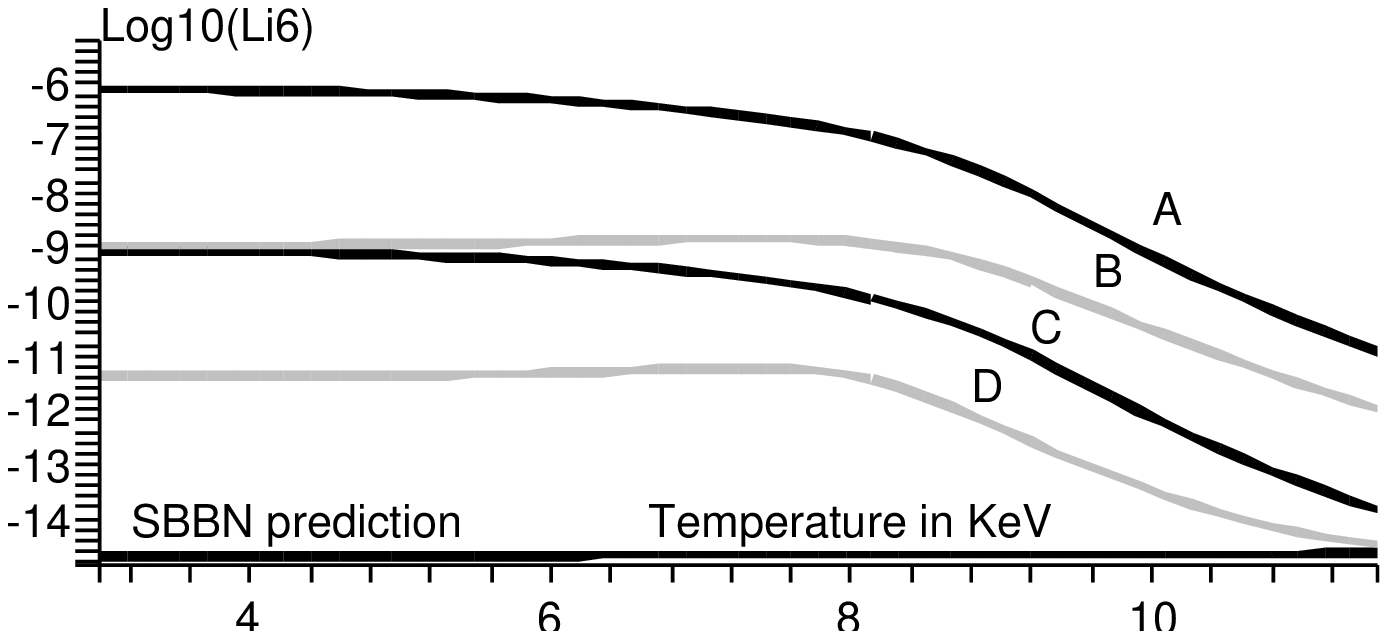}}
\vspace{-.3cm}
 \caption{\footnotesize Log$_{10}(\lisxm)$ plotted as a function of $T$ in KeV for different choices of 
$\tau$ and $Y_X$: A. $\tau=\infty$ and $Y_X=10^{-2}$, B. $\tau=4\times 10^3$\,s and $Y_X=10^{-2}$, 
C. $\tau=\infty$ and $Y_X=10^{-5}$, D. $\tau=4\times 10^3$\,s  and $Y_X=10^{-5}$. }
\vspace{-0.5cm}
\label{f2} 
\end{figure}
Using the Gamow energy that accounts for the charge screening in 
(\hef\xm) and the change of the reduced mass,
\be
E_{\rm SBBN}^{\rm Gamow} = 5249~{\rm KeV}~\to~E_{\rm CBBN}^{\rm Gamow} = 1973~{\rm KeV},
\ee
we calculate the thermally averaged cross section 
for the CBBN processes at small temperatures $T\sim 10$ KeV 
employing a saddle point approximation:
\ba
\langle \sigma_{\rm C}v \rangle \simeq 
2{\rm bn}\times T^{-2/3}\exp(-23.7T^{-1/3})
\nonumber\\\label{sigmav}
= 1.8\times 10^9\times T_9^{-2/3}\exp(-5.37T_9^{-1/3}).
\ea
In the second line, the cross section is converted to customary units of 
$N_A^{-1}$cm$^3$s$^{-1}$g$^{-1}$, and $T_9$ is temperature in units of $10^9$ K. 
CBBN rate (\ref{sigmav}) is to be compared with the SBBN expression \cite{kawano},
\be
\langle \sigma_{\rm S}v \rangle \simeq 30T_9^{-2/3}\exp(-7.435T_9^{-1/3}),
\label{sigmaSBBN}
\ee
and the enhancement of eight orders of magnitude for CBBN is traced back directly to (\ref{Snumeric}).
Although tremendously enhanced relative to (\ref{sigmaSBBN}), the CBBN rate (\ref{sigmav}) 
is by no means larger than the usual photonless reaction rates known to SBBN. 

{\em Catalytic enhancement of {\rm \lisx\,} at $6\simlt T\simlt 12$\,{\rm KeV}.} 
Armed with the rate (\ref{sigmav}), we write down the evolution equation for \lisx\
abundance at $T\simlt 12$ KeV:
\begin{eqnarray}
-HT\fr{d\lisxm}{dT} = {\rm D}( n_{\rm BS}\langle \sigma_{\rm C}v \rangle 
+n_{\rm He}\langle \sigma_{\rm S}v \rangle )
-\lisxm ~n_p \langle \sigma_{\rm p}v \rangle. 
\label{Beq}
\end{eqnarray}
In this formula, D and \lisx\, are the Hydrogen-normalized abundances of these elements,
$n_p$ and $H$ are the temperature-dependent concentration of free protons and the Hubble rate, 
and $\langle \sigma_{\rm p}v \rangle$
is the rate for  \lisx+$p\to$\het+\hef\, responsible for the destruction 
of \lisx\, \cite{kawano}. We use the output of the SBBN code for 
D, \hef\, as function of $T$ and solve the CBBN subset of equations (\ref{rec}) and (\ref{Beq})
numerically.
In fact, since $n_{\rm BS}(T)$ significantly differs from zero only 
below 9 KeV, the variation of \hef\, and D abundances is negligibly small compared to the 
freeze-out values (D$=2.4\times 10^{-5}$, $Y_p = 0.25$).  
Several of the numerical solutions to this equation for different 
input values of $Y_X$ and $\tau$ are plotted in Fig. 2. The increase in time of \lisx\,
below 10 KeV is a direct consequence of CBBN. 
One can clearly see that the with $Y_X\sim O(0.01)$ one 
could convert up to a few \% of D to \lisx. 
Taking the limit of large lifetime ($\tau > 10^5$s), and using the limiting abundance of 
primordial \lisx\, of $2\times 10^{-11}$ \cite{cefo}, we get a remarkable sensitivity to \xm:
\be
Y_X \simlt 3 \times 10^{-7} \longrightarrow n_{X^-}/s\simlt 2.5\times 10^{-17},
\label{constraint}
\ee
A scan over the lifetime parameter produces the exclusion boundary on 
the $(\tau, \log_{10}Y_X)$ plane plotted in Fig. 3. Also shown is the natural range 
for $Y_X$ calculated using the standard freeze-out formula with the input annihilation cross section
in the range of $\sigma_{\rm ann} v = (1.-100.)\times 2\pi\alpha^2m_X^{-2}$ and $m_X=(0.1-1)$TeV. Taken literally,
the \lisx\, overproduction constrains the heavy charged particles to the range of lifetimes 
\be
\tau \simlt 5\times 10^{3}\,{\rm s},
\label{limitontau}
\ee
unless $\sigma_{\rm ann} v$ is tuned to larger values. In fact, both results, (\ref{constraint}) and (\ref{limitontau}), 
should be interpreted as limits of sensitivity to  $(\tau, Y_X)$ via the CBBN enhancement. 
It is true that excessive energy injection at later time may reduce the amount of \lisx\ generated in CBBN, 
and as a consequence relax the constraints of Eq. (\ref{constraint}) \cite{KR}. The combined analysis of two effects,
the CBBN enhancement of \lisx\ and impact of energy release on elemental abundances, has to be done by 
combining the result of this work and previous papers on unstable particles in BBN \cite{cefo,kkm,jj}. However, 
such analysis is necessarily model dependent, and falls outside the scope of the present Letter. In many models, 
however, the decay of \xm\ is not involving huge energy release, and then both (\ref{constraint}) and
(\ref{limitontau}) hold unconditionally.

CBBN
points towards an intriguing possibility for alleviating the lithium problem. 
As is evident from Fig.~3, lifetimes of \xm\, between 2000 and 5000 seconds 
can create a primordial source for \lisx\, thus removing 
some tension between observations of \lisx\, in low metallicity systems 
and explanation of its production using cosmic rays \cite{jj,keith_french}. 
But more importantly, such lifetimes of \xm\, can lead to an overall reduction of 
\lisv\, abundance. At the WMAP-determined baryon density, most of \lisv\, is produced via \bes. 
At $T\sim 30$ KeV and $\tau\sim (2-5)\times 10^3$\,s \xm\, is still present in sufficient quantities
to trigger the formation of (\bes$X$). Once formed, (\bes$X$) facilitates several reaction rates 
that destroy \bes. The decay of \xm\, within the bound state leads to further depletion of 
\bes\, and ultimately of \lisv. Moreover, if the decay of \xm occurs via a virtual $W$-boson, there exists
 a fast electroweak capture process of \xm\, on \bes, which also leads to an overall depletion of 
\lisv. 

Usually, a serious obstacle on the way to solve Li problem is 
the abundance of D and \het/D ratio \cite{cefo,kkm,jj} affected through large energy release, 
but we emphasize that CBNN mechanism itself is inconsequential for 
lighter elements such as D or \het\, 
as the abundance of {\em any} bound states of \xm relative to baryons 
is not larger than $10^{-6}$ once the \lisx\, bounds are satisfied. 

\begin{figure}[htbp]
\centerline{\includegraphics[width=9.5 cm]{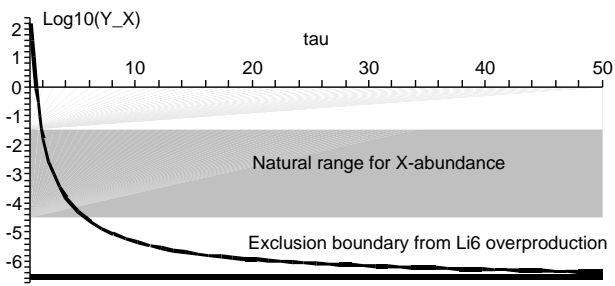}}
\vspace{-.5cm}
 \caption{\footnotesize Constraint on initial abundance of \xm.
$\log_{10}(Y_X)$ is plotted against $\tau$ in units of $10^3$ s. 
The thick horizontal line is the asymptotic exclusion boundary (\ref{constraint}).}
\label{f3} 
\end{figure}

{\em Particle physics applications.} 
There are two generic possibilities
for a long-lived $X^-$: either the decay of \xm\, is suppressed by tiny couplings, 
or it decays to another neutral relic state $X^0$ with small energy release. Schematically, these possibilities
can be represented by
\begin{eqnarray}
\label{types}
&&{\rm Type~I}: ~ X^-\to {\rm SM}^-[X^0], ~~\Delta E\sim M_X,
\\
&&{\rm Type~II}: ~ X^-\to X^0 + e^-[\nu];~~\Delta E\simlt {\rm few~MeV},
\nonumber
\end{eqnarray}
where the first type corresponds to a decay into 
charged SM state(s) with or without any neutral non-SM
relics, and the second type is \xm decaying to another 
neutral relic plus an electron and possibly some neutrino states
with a small energy release. 
SUSY models have scenarios of both types. An example of type I is the 
charged slepton ({\em e.g.} stau) decaying into gravitinos or decaying due to $R$-parity violation
without leaving any SUSY remnants. 
Since the amount of energy released in the decay of \xm\, 
of type I is large, we cannot restrict this scenario only to the CBBN-enhanced output of \lisx\, as 
non-thermal BBN may provide equally important changes in the elemental abundance. In case of very late decays 
the constraints coming from {\em e.g.} diffuse $\gamma$-ray and microwave backgrounds can become important. 

Prior to this work, models of type II
were believed to be of no consequence for BBN. Type II models are not involving significant energy release, 
and therefore our results based on CBBN enhancement, Eqs. (\ref{constraint}) 
and (\ref{limitontau}), become strict limits rather than levels of sensitivity. 
Examples of type II include an important case of near-degenerate states 
of lightest (LSP) and next-to-lightest (NLSP) SUSY particles, 
neutralino-chargino or stau-neutralino. Using recent  calculation Ref. \cite{longlive} of stau lifetime, 
we can translate (\ref{limitontau}) into the limit on {\em  minimal} mass splitting 
in the stau-neutralino system, $m_{\tilde \tau} - m_{\chi} > 70~{\rm MeV}$.
Along the same lines, one can derive constraints on the minimal mass splitting between 
neutral and charged particles at the first excited Kaluza-Klein level in models 
with ``universal" extra dimensions.



To conclude, thermal catalysis of BBN reactions represents the novel idea of how particle 
physics can alter the predictions for light elemental abundances. The catalysis of \lisx\, via the 
production of (\hef\xm) bound states is explicitly demonstrated, and 
the enormous enhancement in the catalyzed channel furnishes the sensitivity to \xm\, normalized on entropy 
at $3\times 10^{-17}$ level. 
I would like to thank  C. Bird and K. Koopmans for 
assistance in some calculations and useful discussions, and acknowledge 
useful conversations with Drs. R. Allahverdi, R. Cyburt, K. Olive, 
A. Ritz and Y. Santoso.

\end{document}